\documentclass[final,3p,times,12pt]{elsarticle}

\usepackage{amssymb}
\usepackage{multirow}
\usepackage{float}
\usepackage{graphicx}
\usepackage{adjustbox}
\journal{Computational Materials Science}

\begin{document}

\begin{frontmatter}

\title{First-principles study of structural disorder, site preference, chemical bonding and transport properties of Mg-doped tetrahedrite}

\author[kckizw]{Krzysztof Kapera\corref{cor}}
\cortext[cor]{Corresponding author}
\ead{kaperak@agh.edu.pl}
\author[kckizw]{Andrzej Koleżyński}
\ead{kolezyn@agh.edu.pl}
 \affiliation[kckizw]{organization={Department of Silicate Chemistry and Macromolecular Compounds, Faculty of Materials Science and Ceramics, AGH University of Science and Technology},
             addressline={al. A. Mickiewicza 30},
             city={Cracow},
             postcode={30-059},
             country={Poland}}

\begin{abstract}
Tetrahedrite-based ($\textrm{Cu}_{12}\textrm{Sb}_{4}\textrm{S}_{13}$) materials are candidates for good thermoelectric generators due to their intrinsic, very low thermal conductivity and high power factor. One of the current limitations is virtual absence of tetrahedrites exhibiting n--type conductivity. In this work, first-principles calculations are carried out to study Mg-doped tetrahedrite, $\textrm{Mg}_{x}\textrm{Cu}_{12}\textrm{Sb}_{4}\textrm{S}_{13}$ with possibility of predicting n--type material in mind. Different concentrations and modifications of the structure are investigated for their formation energies, preferred site occupation and change in local environment around dopants. Mg atoms tend to occupy 6b site, while introduced excess Cu prefers 24g site. Introduction of elements in those sites display different effect on nearby rattling Cu(2) atom. Topological analysis shows that tetrahedrite exhibits ionic, closed-shell bonds with some degree of covalency. Majority of the bonds weakens with increasing content of Mg; structure becomes increasingly less stable, which is also expressed by global instability and bond strain indexes. Achieving n--type conductivity was predicted by the calculations for structures with $x>1.0$, however increasing enthalpy of formation and lack of stability might suggest limit of solubility and difficulties in obtaining those experimentally.
\end{abstract}

\begin{keyword}
thermoelectric material \sep ab initio \sep DFT calculations \sep tetrahedrite \sep WIEN2k \sep Critic2
\end{keyword}

\end{frontmatter}

\section{Introduction}
\label{introduction}

Thermoelectric materials (TE) are environmentally friendly materials that enable direct and reversible conversion of temperature difference to electrical energy. Thermoelectric devices have great potential in waste heat recovery, however their usage is limited by poor performance. The performance of TE is expressed using dimensionless figure of merit, $zT=S^2 \sigma T / \kappa$, where $S$ is Seebeck coefficient, $\sigma$ is electrical conductivity, $T$ is absolute temperature and $\kappa$ is thermal conductivity. It is worth noting that optimization of $zT$ is not easy. Since thermal conductivity includes electronic $\kappa_E$ and lattice $\kappa_L$ contributions ($\kappa=\kappa_E+\kappa_L$), $zT$ is inherently correlated with electrical conductivity, as stated in Wiedemann-Franz law \cite{Snyder-2008}. In addition, the Pisarenko relation \cite{dresselhaus2013overview} limits simultaneous increasement of $S$ and $\sigma$.

TE are generally thermal insulators and electrical conductors. Glen A. Slack proposed a “phonon glass and an electron single crystal” (PGEC) concept \cite{slack1995CRC}, in which material’s phonon mean free path is disrupted without affecting its electron transport properties. To lower heat conducted by crystalline lattice, one can introduce scattering centers of different sizes which work on different parts of acoustic phonon spectra. There can be point defects (e.g. in Si--Ge alloy \cite{zhu2009increased}), “rattlers” (found in more complex structures such as clathrates, partially filled skutterudites as well as tetrahedrites \cite{takabatake2014phonon}), or even molecules and nanocrystals \cite{sales2007critical}.

Tetrahedrite, $\textrm{Cu}_{12}\textrm{Sb}_{4}\textrm{S}_{13}$, is a p--type conductor and has been studied as promising thermoelectric materials for a few reasons. It is a naturally occurring earth abundant mineral, composed of light, non-toxic elements. Its crystal structure is quite complex, with 58 atoms per unit cell, which also contributes to intrinsically low thermal conductivity. It has been reported that tetrahedrite co-doped with Zn and Ni exceeds $zT=1$ above 723 K and exhibits lattice thermal conductivity of $\sim$0.45 W m\textsuperscript{-1} K\textsuperscript{-1} in wide temperature range 300--700 K (which is similar to amorphous structures) \cite{doi:10.1021/cm502570b}.

Tetrahedrite’s crystal structure has been known since Weunsch \cite{wuensch1964crystal}. It crystallizes in regular system with $I\bar{4}3m$ space group, while atoms occupy five different Wyckoff positions: Cu(1) site, at Wyckoff position 12d (atomic coordinates $1/4, 1/2, 0$); Cu(2) site, at Wyckoff position 12e (atomic coordinates $x, 0, 0$); Sb(3) site, at Wyckoff position 8c (atomic coordinates $x, x, x$); S(1) site, at Wyckoff position 24g (atomic coordinates $x, x, z$); S(2) site, at Wyckoff position 2a (atomic coordinates $0, 0, 0$).

In this work while talking about structural voids, $\square$(x) symbol will be used, with “x” denoting Wyckoff position. Tetrahedrite’s chemical formula can be stochiometrically written as Cu(1)\textsubscript{6}Cu(2)\textsubscript{6}Sb(3)\textsubscript{4}S(1)\textsubscript{12}S(2). Tetrahedrite exhibits so-called valency-imposed double-site occupancy \cite{hatert2008ima}. Out of six cations occupying Cu(1) site, four of them are Cu\textsuperscript{+} and two are Cu\textsuperscript{2+} (with random distribution). Thus, charge-wise, tetrahedrite’s formulae can be given as $\textrm{Cu(1)}_{2}^{2+} \textrm{Cu(1)}_{4}^{+} \textrm{Cu(2)}_{6}^{+} \textrm{Sb(3)}_{4}^{3+} \textrm{S(1)}_{12}^{2-} \textrm{S(2)}^{2-}$.

\begin{figure}[H]
  \includegraphics[width=\linewidth]{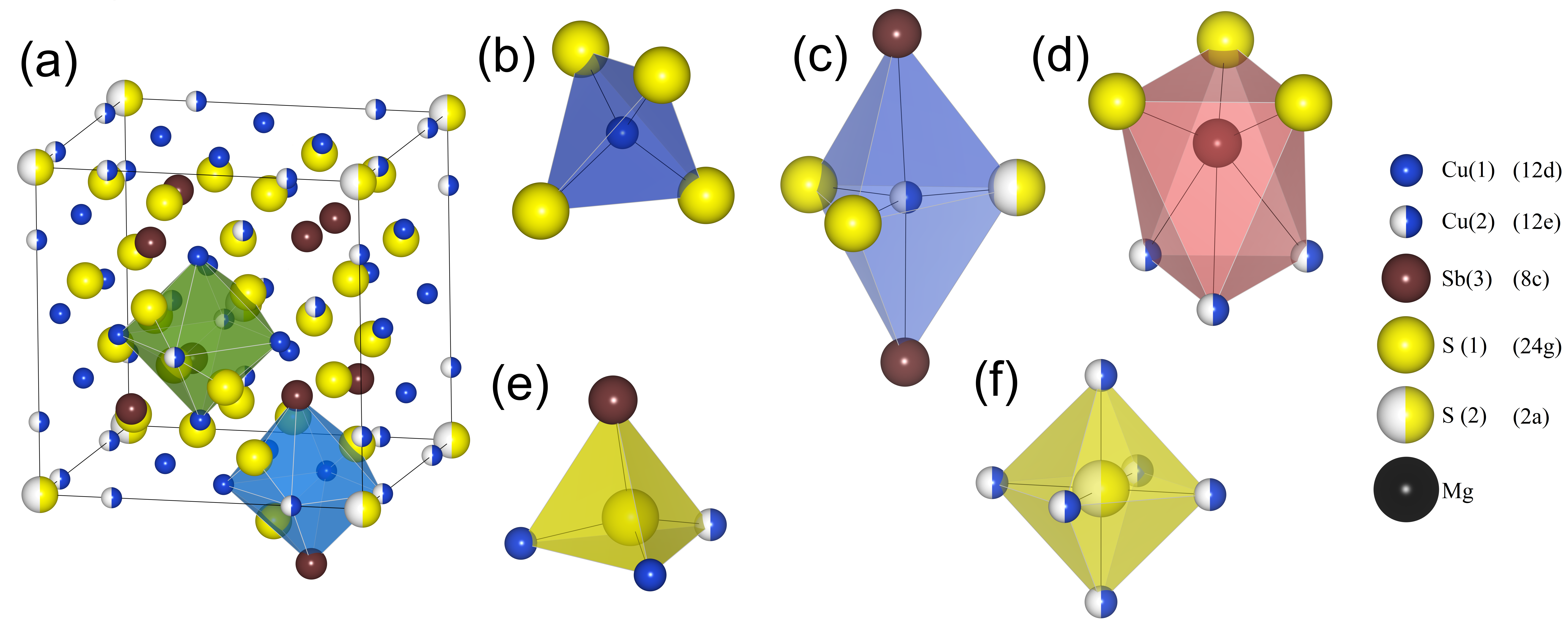}
  \caption{\textbf{(a)} Tetrahedrite's unit cell with marked structural voids with Wyckoff position 6b (green polyhedron) and 24g (blue polyhedron), \textbf{(b-f)} structural units present in tetrahedrite. Magnesium atom in the legend is there for a future reference. This color scheme of atoms will be present throughout this paper.}
  \label{fig:fig1}
\end{figure}

Cu(1) atoms (\textbf{Figure \ref{fig:fig1}}b) are surrounded by four S(1) atoms, forming a tetrahedron. Cu(2) atoms (Figure \ref{fig:fig1}c) are bonded to three planarly distributed sulfur atoms, with two antimony atoms on each side of the {CuS}\textsubscript{3} plane, forming trigonal bypiramid Cu[Sb\textsubscript{2}S\textsubscript{3}]. Sb(3) atoms (Figure \ref{fig:fig1}d) are bonded to three S(1) atoms, forming trigonal pyramid. Sb atoms have free 5sp\textsuperscript{3} electron pair directed toward three Cu(2) atoms. Interaction between this free electron pair and copper atoms will be discussed later. For now, if we were to believe that Sb and Cu(2) atoms form a bond, there would be trigonal antiprism (S\textsubscript{3}SbCu\textsubscript{3}) coordination for antimony. S(1) atoms (Figure \ref{fig:fig1}e) are forming distorted tetrahedron with two Cu(1), one Cu(2) and one Sb(3) at the corners. Finally, S(2) atoms (Figure \ref{fig:fig1}f) have octahedral coordination (six Cu(2) atoms at the corners).

The 12e site exhibits large, anisotropic atomic displacement parameter (ADP) of copper atoms in the direction perpendicular to the Cu(2) -- S\textsubscript{3} plane (U$_\perp=0.13 {\textrm{{\AA}}}^2$ at 300 K \cite{pfitzner1997cu12sb4s13, suekuni2018retreat}). This results in anharmonic out-of-plane rattling modes with low frequency and high amplitude, which can interact with heat carrying phonons – explaining tetrahedrite’s low thermal conductivity. Moreover, tetrahedrite exhibits phonon stiffening upon heating which was again linked to Cu(2) –- Sb interaction \cite{li2016first}. Increase in temperature enables Cu(2) atoms to be closer to Sb atoms, increasing its bonding and resulting in higher vibration energy \cite{lara2014low,bouyrie2015crystal,lai2015bonding}.

More in-depth studies concerning high ADP on 12e site has been done \cite{suekuni2018retreat, lai2015bonding, PhysRevB.102.100302}. Lai et al. \cite{lai2015bonding} believed that high ADP and rattling is due to interaction of Cu(2) atoms with free electron pairs of Sb. This would mean existence of 5-atom atomic cage, with weak and fluctuating bond between Cu(2) and one of the Sb atoms. Umeo et al. \cite{PhysRevB.102.100302} studied effect of pressure on the rattling of tetrahedrite and type-I clathrate. Their results suggested that out-of-plane rattling of Cu(2) atoms is due to chemical pressure. Above the pressure of 2.4 GPa, the anharmonic rattling ceases and Cu(2) site splits into 24g site (Cu22), out of S\textsubscript{3} triangle. Applying lower pressure (up to 0.7 GPa) resulted in decrease of rattling energy. In addition, comparing Cu(2) –- Sb(3) distance with rattling energy suggests only slight impact of antimony free electron pair on the rattling. Area of the S\textsubscript{3} triangle, S\textsubscript{tri} seems to be a dominant factor. Suekuni et al. \cite{suekuni2018retreat} studied crystal structure and phonon dynamics for a series of different compounds from tetrahedrite family. They could not find any correlation between rattling amplitude and “free space” (distance between Cu(2) –- As/Sb), as would be expected for classical caged rattlers. Additionally, inverse correlation between rattling energy, E\textsubscript{R} and out-of-plane ADP, U$_\perp$ has been found.

Undoped tetrahedrite has metallic properties and electrical conductivity of 10\textsuperscript{5} S m\textsuperscript{-1} at 300 K. Pure synthetic $\textrm{Cu}_{12}\textrm{Sb}_{4}\textrm{S}_{13}$ exhibits $zT=0.56$ at 673 K \cite{lu2013high}, however various dopants were introduced by other research teams, leading up to $zT \sim1$ for many cases. There were various attempts to enhance performance of tetrahedrite via doping, co-doping, use of nanostructures, composites or other structural modifications. Known structure modifications include i.a.: dopants introduced in Cu sublattice (Nb \cite{SUN2019835}, Fe \cite{kim2020synthesis, weller2017rapid, nasonova2016role, weller2018observation}, Cd \cite{kumar2016thermoelectric}, Pb \cite{huang2018preparation}, Zn \cite{weller2017rapid, Bouyrie2016, fan2013room}, Ni \cite{weller2017rapid,barbier2015structural,doi:10.1063/1.4789389}, Mn \cite{heo2014enhanced,doi:10.1063/1.4871265,chetty2015thermoelectric,2021chargeMn}, Co \cite{chetty2015thermoelectric}, Cr \cite{kumar2019thermoelectric}, Mg \cite{knivzek2019hbox,levinsky2019thermoelectric}, Ge and Sn \cite{kosaka2017effects}); dopants introduced in Sb sublattice (Bi \cite{kwak2020solid,prem2017thermoelectric}, Te \cite{lu2014effect,bouyrie2015crystal}, Sn \cite{tippireddy2019effect,ahn2021charge}); dopants introduced in S sublattice (Se \cite{tippireddy2018electronic,lu2016phase}); sulfur deficiency \cite{sun2017powder}; excess Cu in structural voids \cite{yan2018high,vaqueiro2017influence} and substitution of Sb for Cu \cite{yang2021influence}. Some of those aforementioned were simultaneous substitutions of more than one element.

In case of most divalent and trivalent transition metals (TM) substituting copper, it has been found that the substitution is preferable at the Cu(1)\textsuperscript{+} 12d site \cite{lu2013high,lopez2021mechanosynthesis,doi:10.1063/1.4789389,chetty2015tetrahedrites}. In some cases however, it can be preferable energetically to replace Cu(1)\textsuperscript{2+} by Cu(1)\textsuperscript{+} during the substitution \cite{makovicky1990role}. It is also possible for simultaneous substitution at both Cu(1) and Cu(2) sites \cite{chetty2015thermoelectric,kumar2019thermoelectric}. Vaqueiro et al. \cite{vaqueiro2017influence} found that copper-rich $\textrm{Cu}_{12+x}\textrm{Sb}_{4}\textrm{S}_{13}$ $(0<x<2)$ phase tends to occupy the Cu(3) site, with Wyckoff position 24g (0.28, 0.28, 0.041). Levinsky et al. \cite{levinsky2019thermoelectric} studied Mg-doped tetrahedrite $\textrm{Mg}_{x}\textrm{Cu}_{12-x}\textrm{Sb}_{4}\textrm{S}_{13}$ and found that it is energetically favorable for Mg atoms to occupy Cu(1) 12d site.

In this work, we use first principle DFT calculations with two objectives in mind. First, to predict which site is favorable energetically for different variants of copper-rich and Mg-doped tetrahedrite. Second, we seek to gain insight on impact of chemical composition, local disorder and character of chemical bonding in tetrahedrite on its fundamental structural properties and stability. 

\section{Theoretical approach and computational details}

First principles calculations were performed via WIEN2k computational package \cite{6165be91106043cc80ee83af3375a9c3}. WIEN2k employs Full-Potential Linearized Augmented Plane Wave (FP-LAPW) approximation, within Density Functional Theory (DFT) formalism, which is suited for solids. For all calculations, following parameters were used: GGA PBEsol as exchange-correlation potential \cite{perdew2008restoring}, R\textsubscript{MT}K\textsubscript{MAX}=8.0, 256 k-points in irreducible Brillouin zone for unmodified structure (and closely matching number of k-points for other structures) and convergence criteria of $\Delta$E= 10\textsuperscript{-5} Ry for energy, $\Delta$q = 10\textsuperscript{-5} e for charge and $\Delta$F = 10\textsuperscript{-1} mRy a\textsubscript{0}\textsuperscript{-1} for forces.\\
First, we created a “vanilla” unmodified structure of tetrahedrite, $\textrm{Cu}_{12}\textrm{Sb}_{4}\textrm{S}_{13}$, then a series of doped/modified structures. Those were:

\begin{itemize}
    \item[\labelitemii] $\textrm{Mg}_{x}\textrm{Cu}_{12}\textrm{Sb}_{4}\textrm{S}_{13}$ ($x=0.5, 1.0, 1.5, 2.0$) in which dopant atoms were introduced into structural voids described by either Wyckoff positions 6b ($0, 1/2, 1/2$) or 24g ($x, x, z$) (total of 8 structures, see \textbf{Figure \ref{fig:fig1}}a for structural voids 6b/24g)
    \item[\labelitemii] $\textrm{Cu}_{12.5}\textrm{Sb}_{4}\textrm{S}_{13}$ in which excess copper was introduced into either 6b or 24g site (total of 2 structures)
    \item[\labelitemii] $\textrm{Mg}_{0.5}\textrm{Cu}_{12}\textrm{Sb}_{4}\textrm{S}_{13}$ in which copper atom from either 12e or 12d Wyckoff position was misplaced into structural void 6b or 24g. Introduced magnesium dopant always took the copper’s usual position (12e or 12d), which effectively results in them exchanging their positions (total of 4 structures)
\end{itemize}

In tetrahedrite, copper atoms tend to be very mobile and are easily displaced. Created structures aim to simulate situations in which atoms are indeed not confined in their original positions. Moreover, structures were optimized (lattice parameters and atomic positions changed with respect to total energy of the system and forces) with their symmetry reduced to $P1$, which allows for unrestricted relaxation.
Next, we performed calculations with \textsc{Critic2} software \cite{otero2014critic2}, which allows for topological analysis of total charge density obtained previously with WIEN2k for relaxed structures. \textsc{Critic2} implements Bader’s atoms in molecules theory (QTAiM). Finally, Brown’s Bond Valence Model (BVM) \cite{brown2009recent} enabled for a bit more global insight into structural properties, including change of valencies, strain factor and global instability index.\\

\section{Results and discussion}

After creating structures, we optimized their lattice parameters and relaxed atomic positions with respect to forces, in a way that minimizes total energy of the structure (without restricting their movement by symmetry and according to forces). Knowing total energy of fully optimized and relaxed structures allowed us to calculate enthalpy of formation, $H_F$. It is defined as energy difference between total energy of given structure and sum of total energies of pure element structures, in their thermodynamically stable forms. For instance, for $\textrm{Mg}_{2}\textrm{Cu}_{12}\textrm{Sb}_{4}\textrm{S}_{13}$:

\begin{equation} \label{eq:1}
   H_F^{{\textrm{Mg}_{2}\textrm{Cu}_{12}\textrm{Sb}_{4}\textrm{S}_{13}}}=E_{tot}^{\textrm{Mg}_{2}\textrm{Cu}_{12}\textrm{Sb}_{4}\textrm{S}_{13}}-12 \cdot E_{tot}^{Cu} - 4 \cdot E_{tot}^{Sb} - 13 \cdot E_{tot}^{S} - 2 \cdot E_{tot}^{Mg}
\end{equation}
\\
where both $H_F$ and $E_{tot}$ are in eV/atom. Structures of pure elements were also created, optimized and relaxed.

\begin{figure}[H]
  \includegraphics[width=\linewidth]{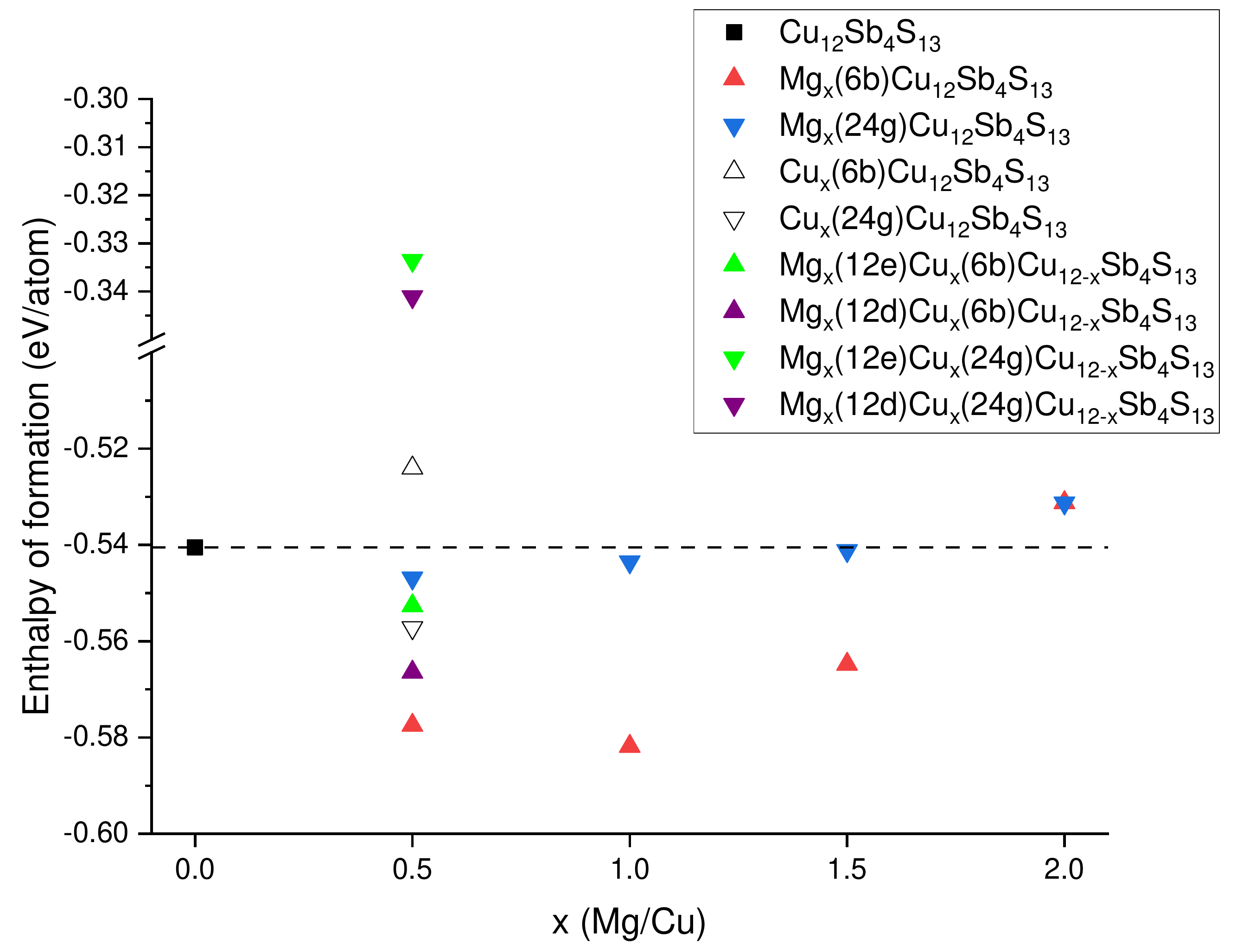}
  \caption{Enthalpies of formation for optimized and relaxed structures, calculated based on Equation \ref{eq:1}}
  \label{fig:fig2}
\end{figure}

Results of enthalpy of formation (see \textbf{Figure \ref{fig:fig2}}) can point us to few conclusions. First of all, for $\textrm{Mg}_{x}\textrm{Cu}_{12}\textrm{Sb}_{4}\textrm{S}_{13}$, Mg atoms seems to prefer $\square$(6b) over $\square$(24g). In both cases, one can see clearly that the lowest enthalpy of formation occurs for slightly doped structures ($x=0.5, 1.0$). Structures with higher concentration of Mg are less stable, which might in fact suggest limit of solubility. However, predicting formation of foreign phases/precipitates such as Cu\textsubscript{3}SbS\textsubscript{4}, CuSbS\textsubscript{2} or MgS is beyond the scope of this work.
\\
As mentioned before, Vaqueiro et al. studied copper-rich tetrahedrite, and reported that additional Cu atoms preferred 24g site. Our results of enthalpy of formation are in agreement with their conclusions. Enthalpy of formation of our $\textrm{Mg}_{0.5}\textrm{Cu}_{12}\textrm{Sb}_{4}\textrm{S}_{13}$ structures with Mg occupying copper’s initial position show that Mg in 12d Wyckoff position is preferable (over 12e). In this case however, it is preferable for Cu atoms to occupy $\square$(6b). Overall, results of enthalpy of formation reflects how flexible the atomic positions in tetrahedrite system are.

\begin{figure}[H]
  \includegraphics[width=\linewidth]{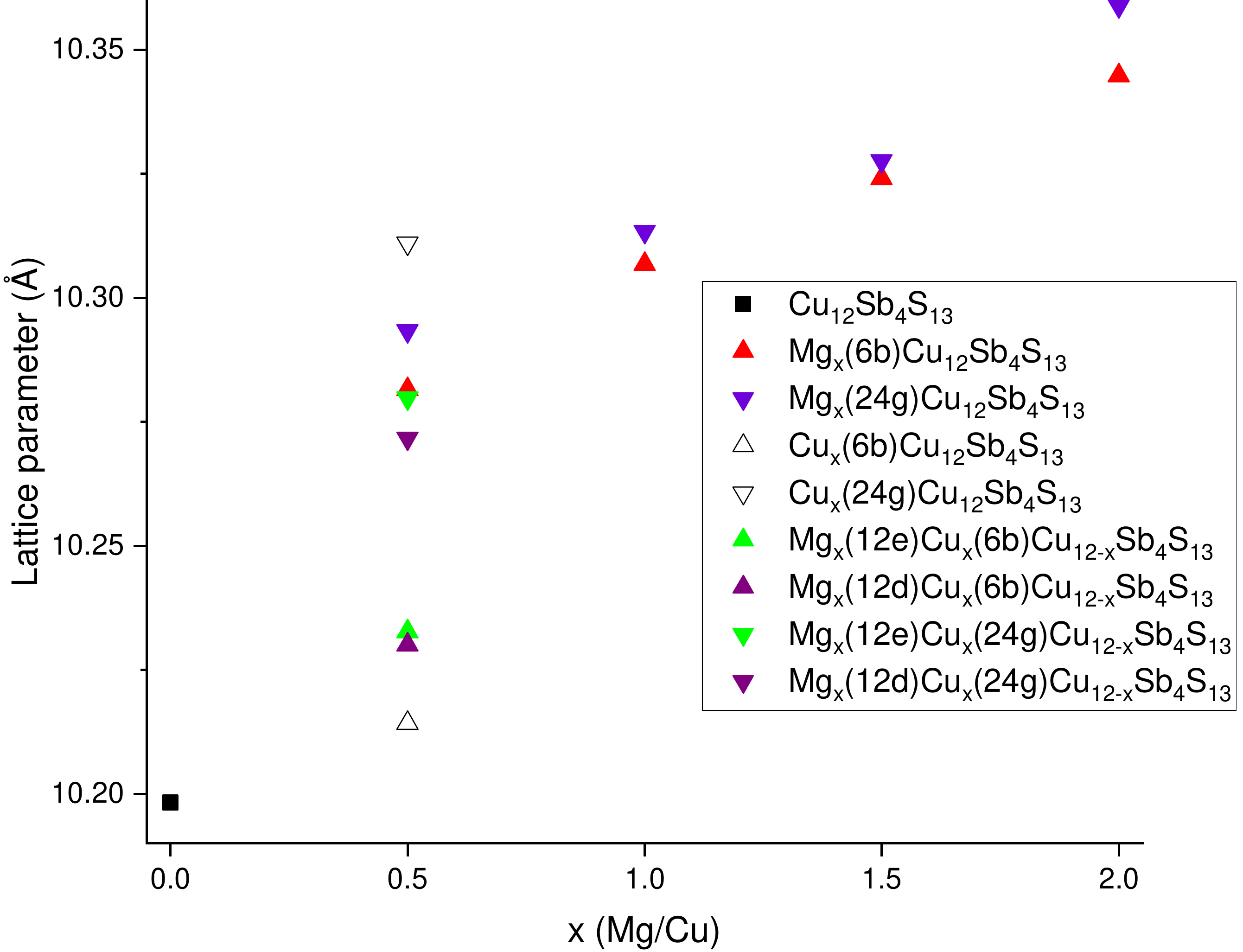}
  \caption{Lattice parameter for optimized tetrahedrite structures obtained by DFT calculations.}
  \label{fig:fig3}
\end{figure}

From \textbf{Figure \ref{fig:fig3}} it is clear that introduction of magnesium/copper into $\square$(24g) results in much higher increase in lattice parameter, than for analogous 6b structures, for all cases. Fitting linear relation yields $0.084x$ $\textrm{{\AA}}$ for 6b and $0.094x$ $\textrm{{\AA}}$ for 24g increase in lattice parameter, which is similar to what has been reported by Levinski et al. \cite{levinsky2019thermoelectric}; approximately 0.1 $\textrm{{\AA}}$ increase per Mg a.p.f.u. (atom per formula unit).

\begin{figure}
  \includegraphics[width=\linewidth]{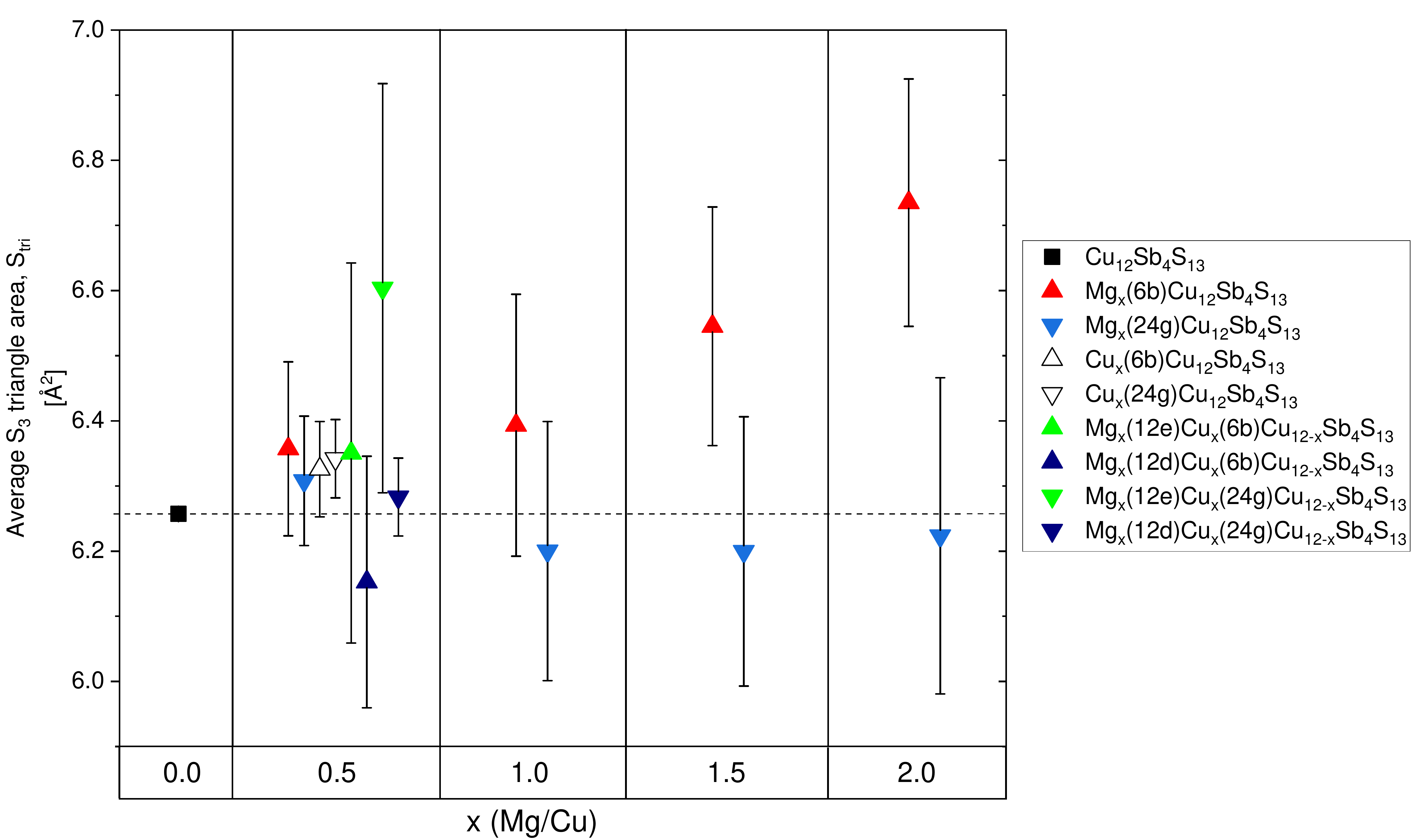}
  \caption{Averaged area of the S\textsubscript{3} triangle (S\textsubscript{tri}) around the Cu(2) 12e sites. Error bars indicate standard deviation.}
  \label{fig:fig4}
\end{figure}

\begin{figure}[H]
  \includegraphics[width=\linewidth]{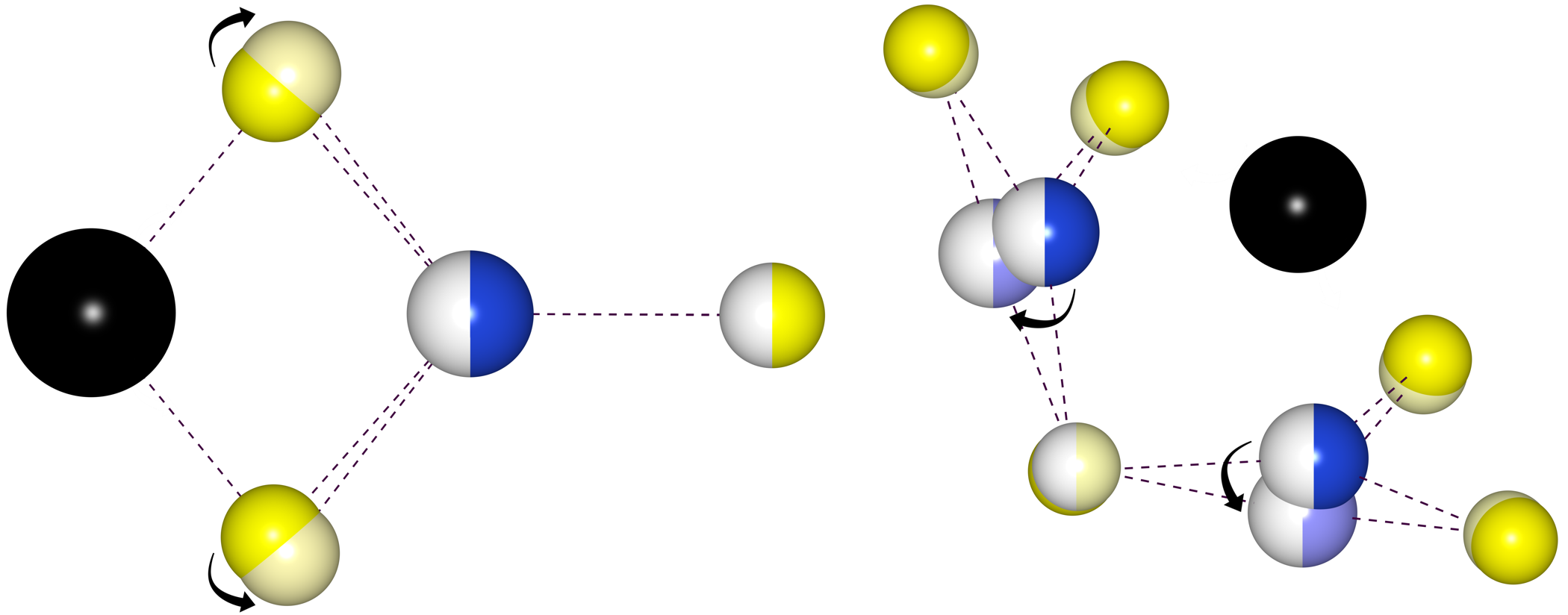}
  \caption{Impact of Mg introduced into 6b (left) and 24g (right) structural void on atomic positions of Mg’s local environment. Unmodified structure (high saturation of colors) and $\textrm{Mg}_{1.0}\textrm{Cu}_{12}\textrm{Sb}_{4}\textrm{S}_{13}$ structures (lower saturation of colors) are superimposed on each other. In both cases, atoms are moving away from introduced Mg.}
  \label{fig:fig5}
\end{figure}

In case of S\textsubscript{3} triangle area (\textbf{Figure \ref{fig:fig4}}), big discrepancy of values for individual structures has been found (5-10\% for most, but up to 18\% for $\textrm{Mg}_{x}\textrm{(12e)}\textrm{Cu}_{x}\textrm{(6b)}\textrm{Cu}_{12-x}\textrm{Sb}_{4}\textrm{S}_{13}$). As mentioned previously, structures for which calculations has been done were allowed to be optimized despite symmetry. Addition of foreign atoms or change in atomic positions often lead to dramatic changes in their local environment.
\\
For a reference, average value of S\textsubscript{tri} for unmodified $\textrm{Cu}_{12}\textrm{Sb}_{4}\textrm{S}_{13}$ is 6.25 $\textrm{{\AA}}^2$. Structures doped with Mg at $\square$(6b) show gradual increase of average S\textsubscript{tri} with Mg concentration. Locally, magnesium atoms at $\square$(6b) are forcing neighbor sulfur atoms to move apart, significantly increasing one side of the triangle (see \textbf{Figure \ref{fig:fig5}}). Depending on the structure, S\textsubscript{tri} for “Mg at 6b” systems can locally reach up to 6.91 $\textrm{{\AA}}^2$. Impact of Cu atom at $\square$(6b) on S\textsubscript{tri} is similar, but much smaller (S\textsubscript{tri} reaches values of up to 6.48 $\textrm{{\AA}}^2$). On the other hand, Mg atoms at $\square$(24g) does not necessarily increase S\textsubscript{tri}. Rather, they displace nearby Cu(2) atoms in the direction perpendicular to S\textsubscript{3} plane, similar to the split-site model; Cu(2) splits into Cu(22) 24g site (see Figure 5). Again, excess Mg atoms seems to have greater impact on their local surrounding than Cu. There are no big differences in atomic positions (not to mention split-site) in the structures in which Cu occupies $\square$(24g). Finally, introduction of Mg atoms into 12e site significantly increases its local S\textsubscript{tri} value (up to 7.57 $\textrm{{\AA}}^2$). It is worth noting that despite increase in lattice parameter for modified structures, some values of S\textsubscript{tri} are still smaller than those for unmodified structure.

\begin{figure}[H]
  \includegraphics[width=\linewidth]{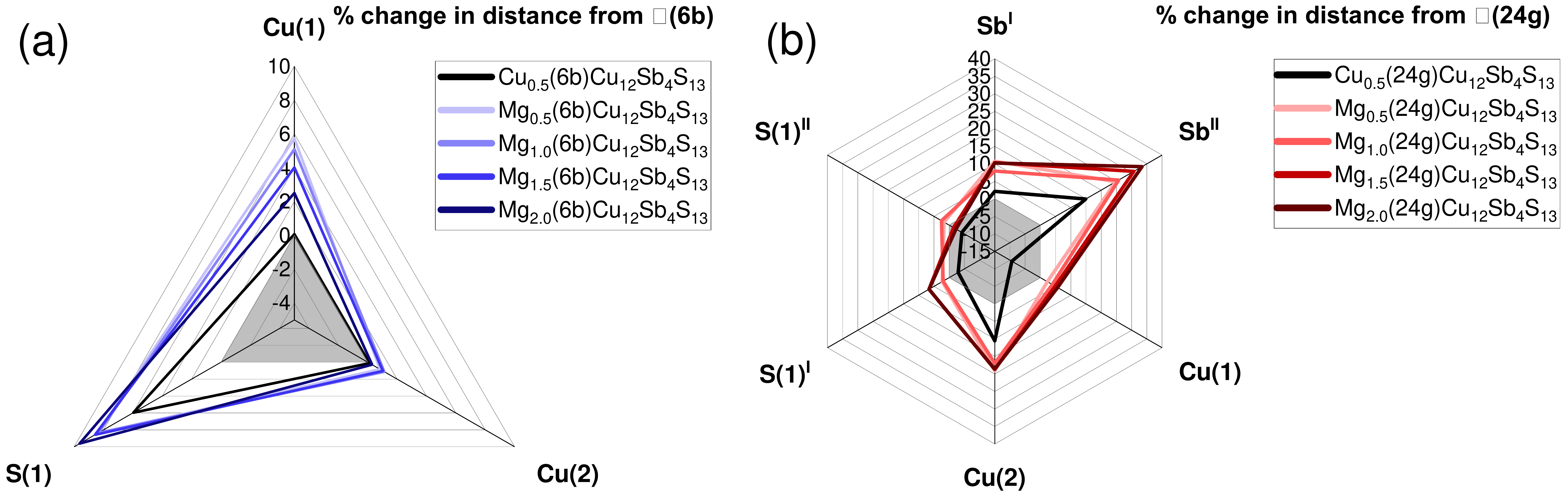}
  \caption{Average change in distance between center of a) $\square$(6b); b) $\square$(24g) and the closest neighbor. Grey area marks distances for unmodified $\textrm{Cu}_{12}\textrm{Sb}_{4}\textrm{S}_{13}$ tetrahedrite structure.}
  \label{fig:fig6}
\end{figure}

\textbf{Figure \ref{fig:fig6}} showcases local impact of introduction of atoms into structural voids by presenting change in distances between center of the $\square$(6b/24g) and neighbor atoms. In the case of 6b systems, there is no increase in the distance between Mg/Cu(6b) and Cu(2) atoms. Introduction of Cu at 6b site also has no effect on the Cu(6b) –- Cu(1) distance. For all of the “6b” systems the most notable change is in Mg/Cu(6b) –- S(1) distances, which increase up to 9.6\%. For the “24g” systems, it should be noted that positions of introduced dopants slightly varied due to relaxation ($x, x, z$ coordinates for 24g site were not fixed). The biggest increase in the distance occurs for Sb atoms, followed by Cu(2); both are substantially larger as compared to 6b systems. Nearby atoms are permitted to move more “freely” and to a higher degree when an additional atom is being introduced into 24g site, due to its lower symmetry.\\

Due to the large volume of raw topological data, most of the results obtained via \textsc{Critic2} software can be found in the Appendix (Tables A.1-4). Results for the tetrahedrite with Mg introduced into $\square$(6b) ($x=0.5$) are presented in \textbf{Table \ref{tab:tab1}}.

\begin{table}[H]
   \caption{Bond length \textbf{R}, eigenvalues of the hessian matrix \boldmath$\lambda_{1-3}$, electron density \boldmath$\rho(r)$, laplacian \boldmath$\nabla^2\rho(r)$, potential energy \boldmath$V[\rho(r)]$, total local energy density \boldmath$H_e[\rho(r)]$, kinetic energy density \boldmath$G[\rho(r)]$, bond’s valency \textbf{Val} and number of given bond type per unit cell \textbf{Mult} for $\textrm{Mg}_{0.5}\textrm{(6b)}\textrm{Cu}_{12}\textrm{Sb}_{4}\textrm{S}_{13}$ tetrahedrite. All values are at the Bond Critical Point (BCP).}
 \label{tab:tab1}
\begin{adjustbox}{max width=\textwidth}
  \begin{tabular}[htbp]{@{}ccccccccccccc@{}}
    \hline
     & & R & $\lambda_1$ & $\lambda_2$& $\lambda_3$ & $\rho(r)$ & $\nabla^2\rho(r)$ & $V[\rho(r)]$ & $H_e[\rho(r)]$ & $G[\rho(r)]$ & Val & Multi\\
      & & [{\AA}] & [au$\cdot10^{-2}$] & [au$\cdot10^{-2}$] & [au$\cdot10^{-2}$] & [au$\cdot10^{-2}$] & [au$\cdot10^{-2}$] & [au] & [au] & [au] &  & \\
    \hline
    \multirow{4}{*}{r\textsubscript{1}} & \multirow{4}{*}{Cu -- Sb} & 3.313 & -0.796 & -0.726	& 3.808 &	1.564 & 2.286 & -0.037 & -0.016 & 0.022 & 0.130 & 4\\
    & & 3.300 & -0.784 & -0.727 & 3.843 & 1.580 & 2.332 & -0.047 & -0.021 & 0.027 & 0.134 & 8\\
    & & 3.300 & -0.807 & -0.759 & 3.891 & 1.590 & 2.325 & -0.040 & -0.017 & 0.023 & 0.134 & 8\\
    & & 3.109 & -1.227 & -1.200 & 5.451 & 2.160 & 3.023 & -0.177 & -0.084 & 0.092 & 0.225 & 4\\
    \hline
    \multirow{4}{*}{r\textsubscript{2}} & \multirow{4}{*}{S(2) -- Cu(2)} & 2.291 & -5.839 & -5.320 & 25.731 & 7.142 & 14.572 & -0.858 & -0.411 & 0.447 & 0.295 & 2\\
    & & 2.282 & -5.952 & -5.509 & 26.291 & 7.267 & 14.830 & -0.871 & -0.417 & 0.454 & 0.302 & 4\\
    & & 2.281 & -6.072 & -5.596 & 26.607 & 7.296 & 14.940 & -0.880 & -0.421 & 0.458 & 0.303 & 2\\
    & & 2.268 & -6.199 & -5.708 & 27.184 & 7.468 & 15.277 & -0.894 & -0.428 & 0.466 & 0.313 & 4\\
    \hline
   \multirow{6}{*}{r\textsubscript{3}} & \multirow{6}{*}{S(1) -- Cu(1)} & 2.283 & -5.877 & -5.596 & 27.026 & 7.433 & 15.552 & -0.807 & -0.384 & 0.423 & 0.234 & 8\\
   & & 2.282 & -5.924 & -5.657 & 27.137 & 7.478 & 15.556 & -0.810 & -0.386 & 0.425 & 0.235 & 8\\
   & & 2.281 & -5.915 & -5.699 & 27.172 & 7.479 & 15.557 & -0.809 & -0.385 & 0.424 & 0.236 & 8\\
   & & 2.274 & -6.033 & -5.709 & 27.495 & 7.588 & 15.753 & -0.825 & -0.393 & 0.432 & 0.240 & 8\\
   & & 2.251 & -6.335 & -6.310 & 29.486 & 7.951 & 16.841 & -0.867 & -0.412 & 0.454 & 0.256 & 8\\
   & & 2.419 & -3.964 & -2.612 & 16.740 & 5.268 & 10.165 & -0.555 & -0.265 & 0.290 & 0.162 & 8\\
   \hline
   \multirow{4}{*}{r\textsubscript{4}} & \multirow{4}{*}{S(1) -- Sb} & 2.459 & -5.696 & -5.487 & 16.598 & 8.057 & 5.415 & -1.029 & -0.508 & 0.521 & 0.854 & 8\\
   & & 2.543 & -4.546 & -4.379 & 13.246 & 6.430 & 4.321 & -0.821 & -0.405 & 0.416 & 0.681 & 4\\
   & & 2.472 & -5.504 & -5.301 & 16.038 & 7.785 & 5.232 & -0.994 & -0.491 & 0.504 & 0.825 & 8\\
   & & 2.469 & -5.632 & -5.424 & 16.409 & 7.965 & 5.353 & -1.017 & -0.502 & 0.515 & 0.832 & 4\\
   \hline
   \multirow{4}{*}{r\textsubscript{4}} & \multirow{4}{*}{S(1) -- Cu(2)} & 2.278 & -5.967 & -5.574 & 27.235 & 7.563 & 15.694 & -0.852 & -0.406 & 0.446 & 0.571 & 4\\
   & & 2.238 & -6.621 & -6.343 & 29.569 & 8.102 & 16.606 & -0.903 & -0.431 & 0.472 & 0.635 & 8\\
   & & 2.238 & -6.606 & -6.366 & 29.597	& 8.141 & 16.625 & -0.905 & -0.431 & 0.473 & 0.637 & 4\\
   & & 2.233 & -6.726 & -6.475 & 29.995 & 8.207 & 16.794 & -0.910 & -0.434 & 0.476 & 0.645 & 8\\
   \hline
   r\textsubscript{6} & S(1)--Mg & 2.349 & -3.199 & -3.109 & 22.870 & 3.863 & 16.561 & -1.013 & -0.486 & 0.527 & 0.500 & 4\\
   \hline
  \end{tabular}
  \end{adjustbox}
\end{table}

\begin{figure}[H]
  \includegraphics[width=\linewidth]{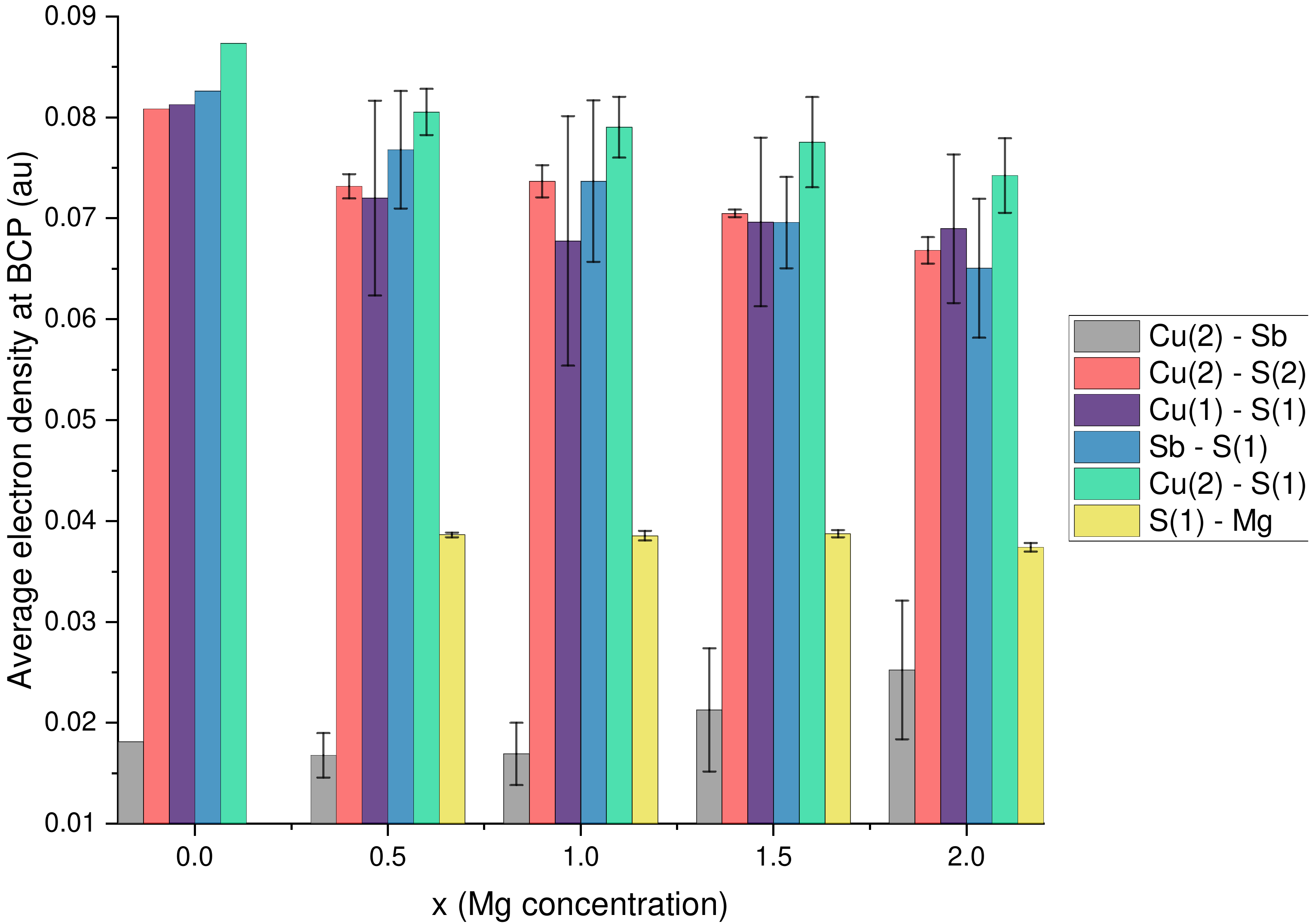}
  \caption{Average values of electron density at BCP for $\textrm{Mg}_{x}\textrm{(6b)}\textrm{Cu}_{12}\textrm{Sb}_{4}\textrm{S}_{13}$ for all types of bonds present. Error bars indicate standard deviation.}
  \label{fig:fig7}
\end{figure}

First of all, our calculations has found a bond critical point between Cu(2) –- Sb, which implies that free electron pairs of antimony are not as free. Instead, they form weak bond with Cu(2) 12e atoms (as depicted in Figure \ref{fig:fig1}). Cu(2) –- Sb interaction was not always certain to be a bond, but has been reported before \cite{yang2020stability}.
\\
Values of electron density at BCP (see Table \ref{tab:tab1}) are small, which suggests that tetrahedrite is predominantly a closed-shell interaction system. Average values of $\rho_{BCP}(r)$ are presented at \textbf{Figure \ref{fig:fig7}}. In general, $\rho_{BCP}(r)$ is strongly correlated with bond order (strength of chemical bond) and bond length. Cu(2) -- Sb bonds are by far the longest and the weakest in tetrahedrite. Introduction of magnesium into $\square$(6b) results in existence of both longer and shorter Cu(2) -– Sb bonds (as compared to unmodified structure), with substantial differences in length. For instance in unmodified tetrahedrite, Cu(2) –- Sb distance is 3.22 $\textrm{{\AA}}$; in $\textrm{Mg}_{1.5}\textrm{(6b)}\textrm{Cu}_{12}\textrm{Sb}_{4}\textrm{S}_{13}$ structure, due to symmetry split, values are 2.97 and 3.33 $\textrm{{\AA}}$. For structures with Mg introduced in $\square$(24g) there is even higher order of distortion, for example in $\textrm{Mg}_{1.0}\textrm{(24g)}\textrm{Cu}_{12}\textrm{Sb}_{4}\textrm{S}_{13}$ structure, Cu(2) -- Sb bonds have values of 2.77 $\textrm{{\AA}}$ and 3.66 $\textrm{{\AA}}$. Despite the increase in $\rho_{BCP}(r)$ standard deviation values for Cu(2) -- Sb, those bonds on average also increase their strength.

Another interesting point to make is about S(1) -- Mg bonds. From Figure 7, it is clear that their values of electron density at BCP are almost constant, no matter the amount of introduced Mg (which is also reflected in similar bond length across structures). On the other hand, all remaining bonds change their electron density at BCP by quite a bit, as indicated by error bars in Figure 7. This would suggest it is more energetically favorable for the structures to adjust position and bond lengths for Cu, S and Sb atoms than Mg. General trend, observed for most bonds, of decreasing $\rho_{BCP}(r)$ and increasing average bond lengths is partially due to increase in unit cell volume with introduction of Mg.

Usually, small and positive values of Laplacian $\nabla^2\rho_{BCP}(r)$, would again indicate that tetrahedrite is a closed-shell system, with local depletion of electron density at BCP. However, Cremer and Kraka \cite{cremer1984chemical} pointed out that characterization of bonds by their $\nabla^2\rho_{BCP}(r)$ is insufficient. They noticed that total local energy density $H_e$ is always negative for covalent bonding (which is the case for tetrahedrite). Finally, Espinoza et al. \cite{espinosa2002weak} proposed a region between pure closed-shell and shared-shell systems, which is described by $1<V[\rho_{BCP}(r)]/G[\rho_{BCP}(r)]<2$ and $(-G[\rho_{BCP}(r)]<H_e[\rho_{BCP}(r)]<0$. It was associated with formation of bonding MO. Based on values of $\nabla^2\rho_{BCP}(r),V,G$ and $H_e$ obtained from topological calculations, tetrahedrite seems to fall under closed-shell, but partially covalent system described above. As proposed by Mori-S\'anchez et al. \cite{mori2002classification}, $f$-index (valence electron density flatness ratio $f=(\rho_{CCP}(r)^{min})⁄(\rho_{BCP}(r)^{max})$ reaches values around 0.047 in all tetrahedrite systems. Global charge-transfer index, $c$ (averaged ratio between topological charge and nominal oxidation state) across tetrahedrite structures fluctuates between $0.37\div0.41$, which again indicates that tetrahedrite is an ionic crystal, with some degree of covalency.

\begin{figure}[H]
  \includegraphics[width=\linewidth]{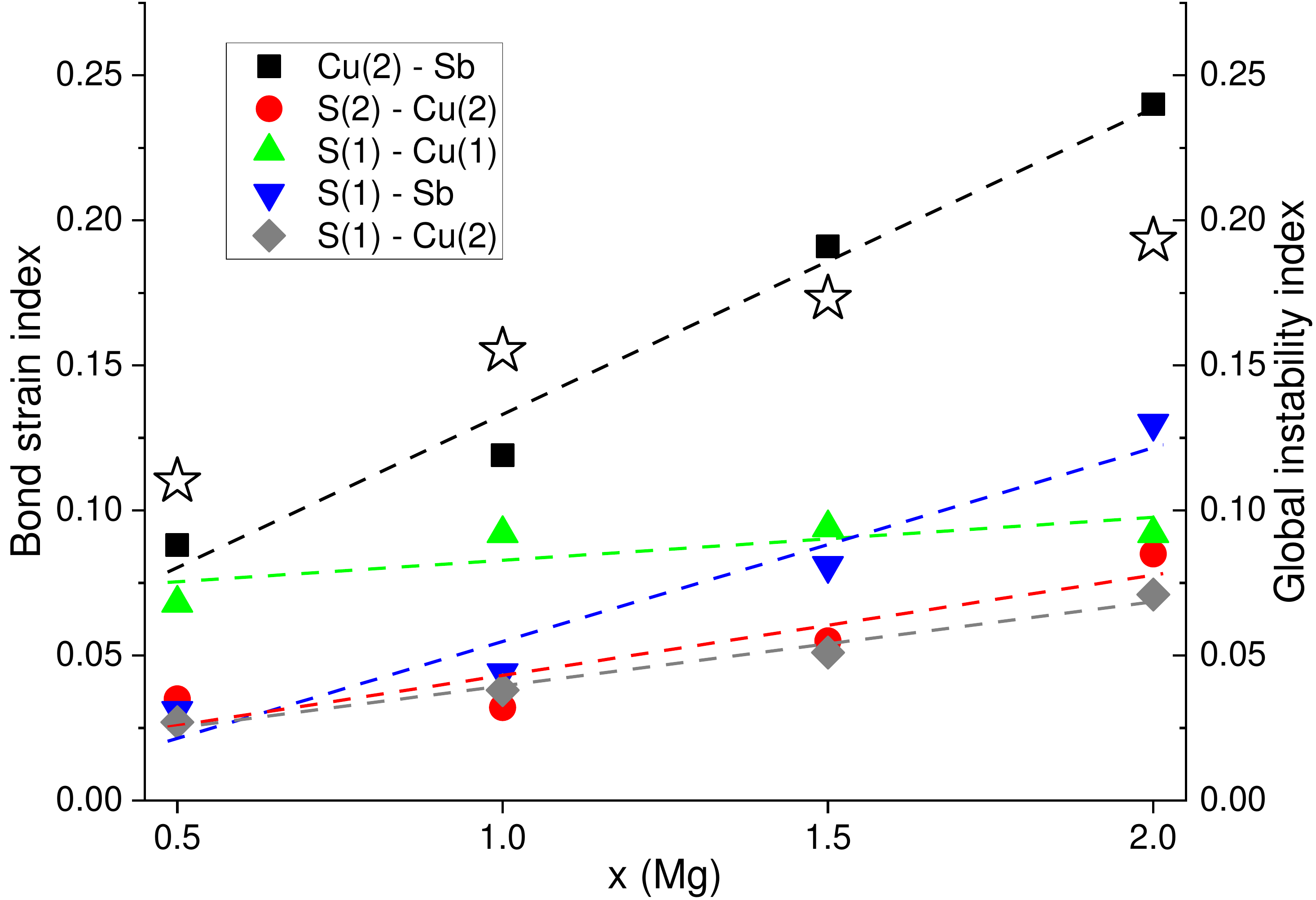}
  \caption{Bond strain index (full symbols) and global instability index (open stars) plots for $\textrm{Mg}_{x}\textrm{(6b)}\textrm{Cu}_{12}\textrm{Sb}_{4}\textrm{S}_{13}$ tetrahedrite structures. Dashed lines for bond strain index are to guide the eye.}
  \label{fig:fig8}
\end{figure}

Bond’s valencies (in Table \ref{tab:tab1}), global instability index, $G$ and bond strain index, $B$ were calculated based on Brown’s Bond Valence Theory \cite{brown2009recent}. \textbf{Figure \ref{fig:fig8}} shows gradual increase in both indexes for $\textrm{Mg}_{x}\textrm{(6b)}\textrm{Cu}_{12}\textrm{Sb}_{4}\textrm{S}_{13}$ tetrahedrite. Even lightly doped structure ($x=0.5$) is strained, however with higher concentration of Mg ($x=1.5, 2.0$) it should be expected that tetrahedrite enters "unstable” range. Especially for higher Mg concentrations, it is mostly Cu(2) –- Sb and S(1) –- Sb bonds that deviate from ideal lengths. This is due to Sb atoms greatly changing their coordinates in the local presence of introduced Mg.

\begin{figure}[H]
  \includegraphics[width=\linewidth]{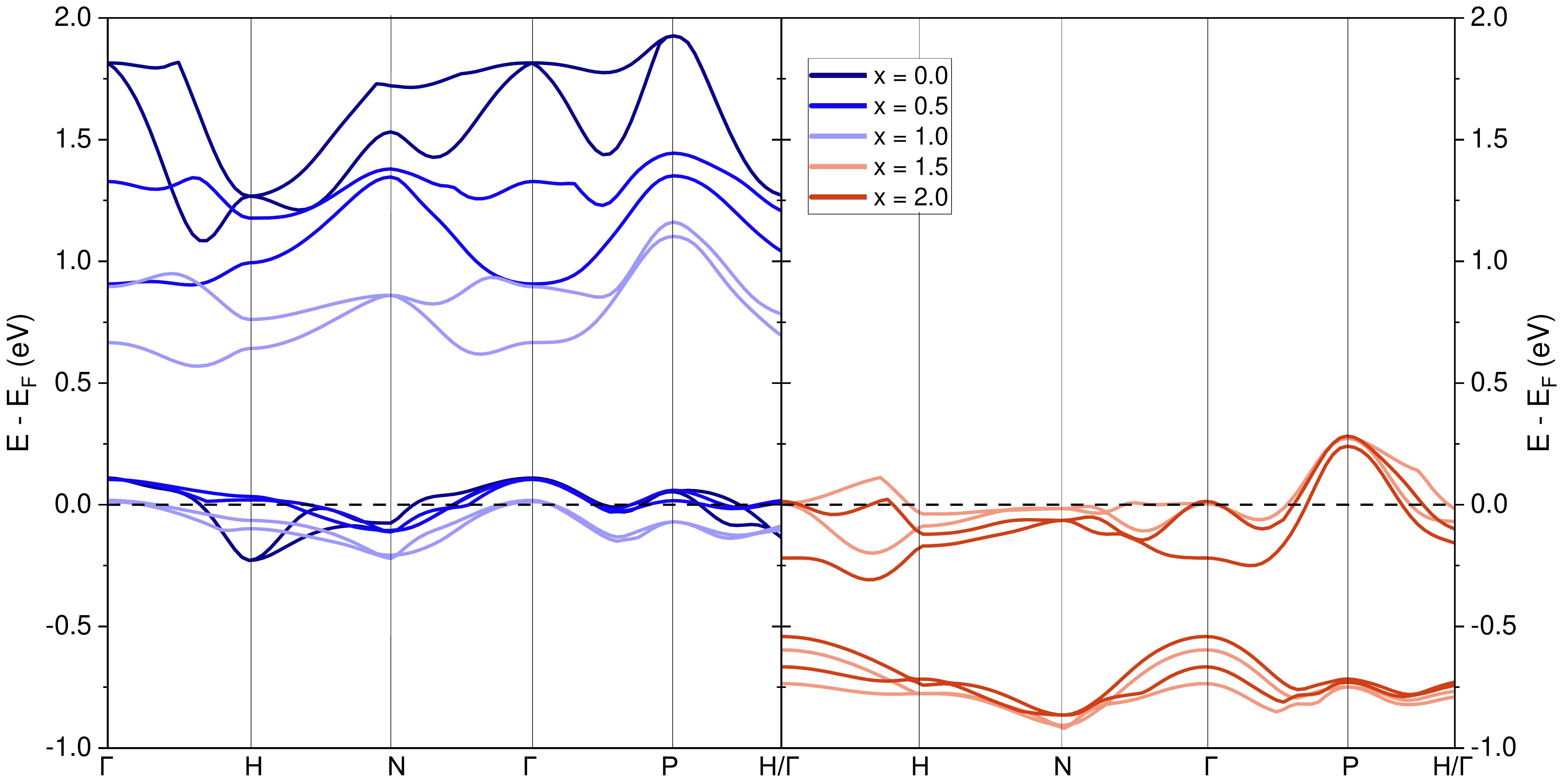}
  \caption{Two highest valence and two lowest conduction bands for $\textrm{Mg}_{x}\textrm{(6b)}\textrm{Cu}_{12}\textrm{Sb}_{4}\textrm{S}_{13}$ tetrahedrite. Whole band structure was not shown for clarity purposes.}
  \label{fig:fig9}
\end{figure}

\begin{figure}[H]
  \includegraphics[width=\linewidth]{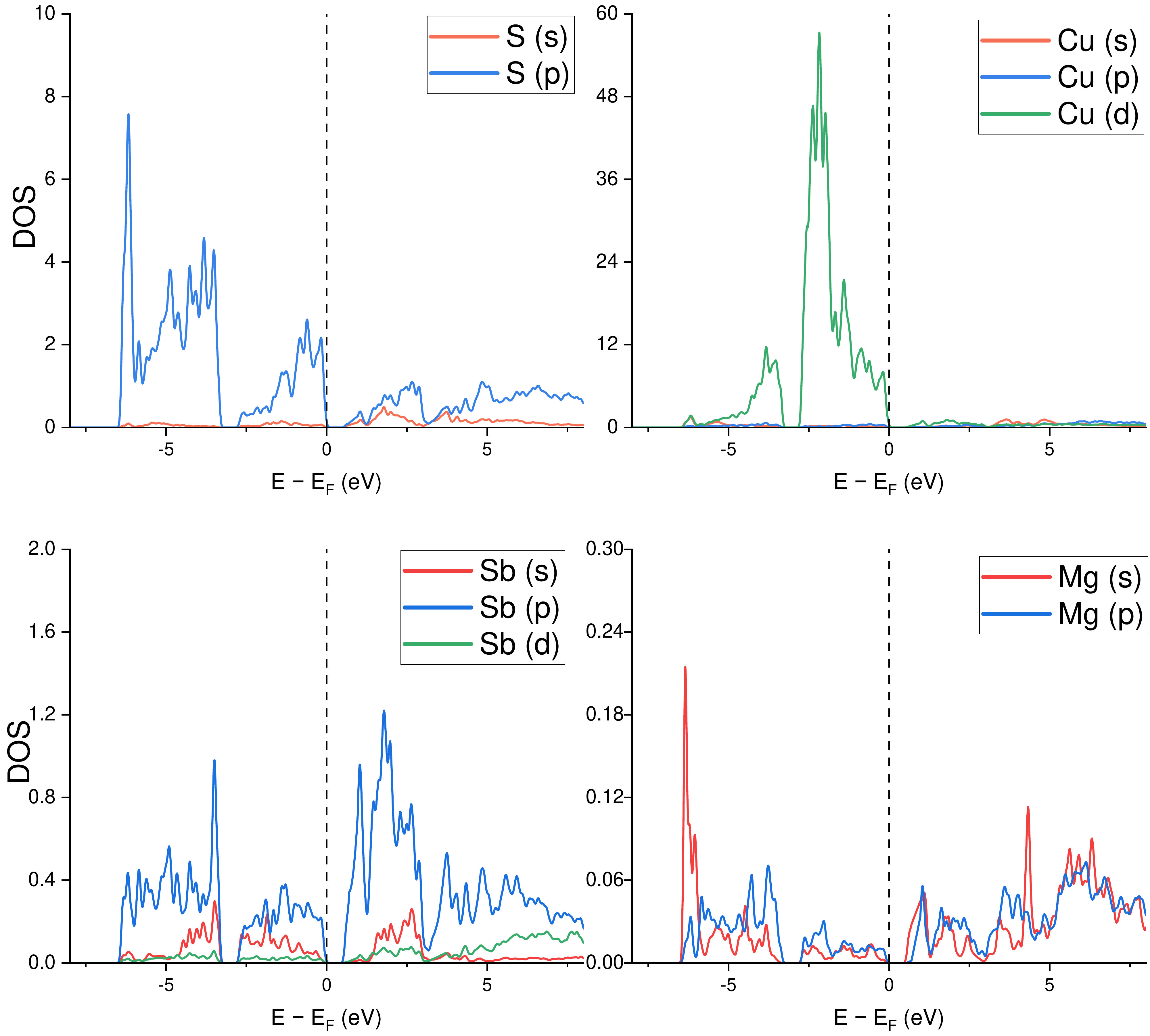}
  \caption{Partial Density of States plots for $\textrm{Mg}_{1}\textrm{(6b)}\textrm{Cu}_{12}\textrm{Sb}_{4}\textrm{S}_{13}$ tetrahedrite. Black dashed line marks Fermi level.}
  \label{fig:fig10}
\end{figure}

\begin{figure}[H]
  \includegraphics[width=\linewidth]{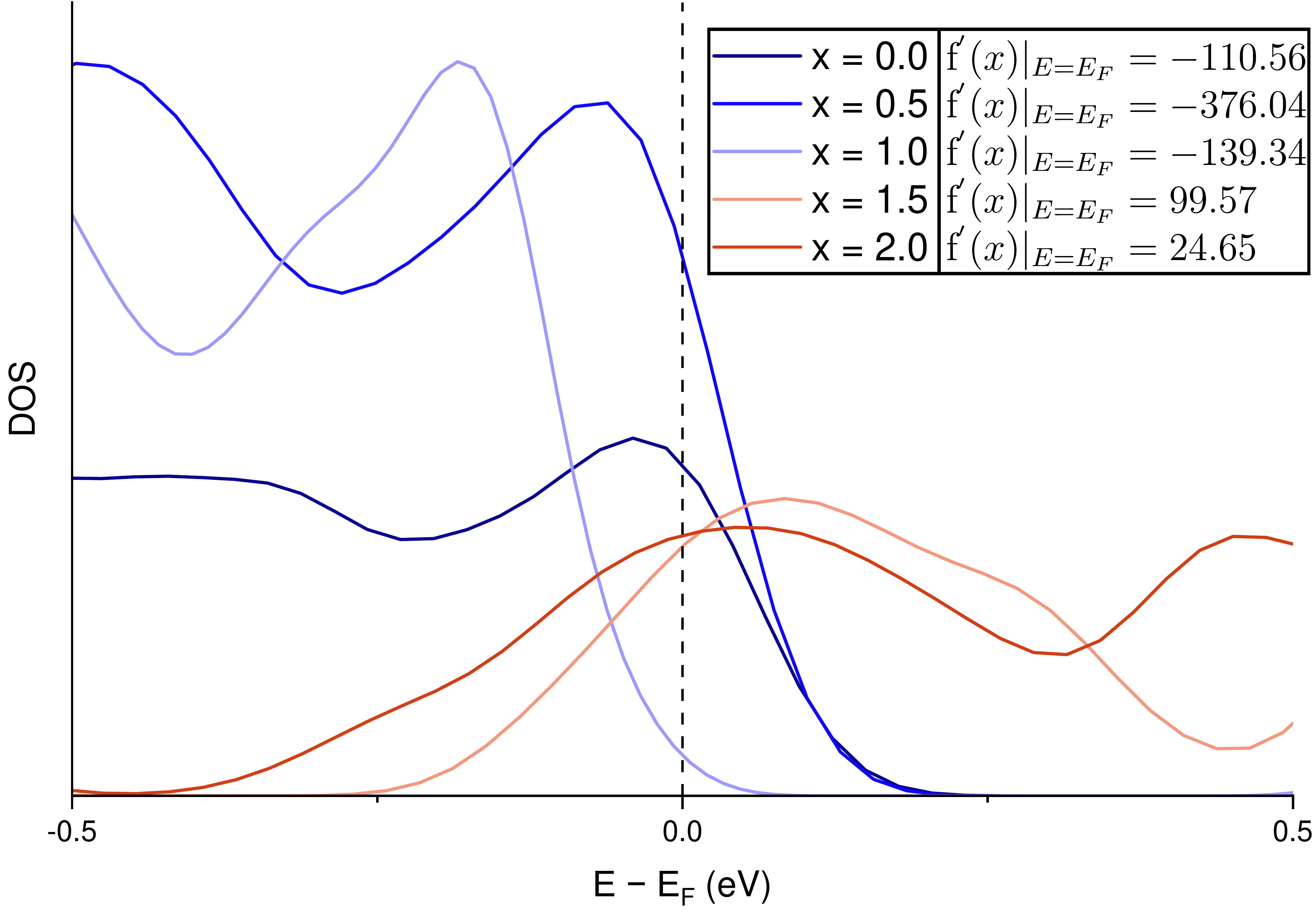}
  \caption{Total Density of States for $\textrm{Mg}_{x}\textrm{(6b)}\textrm{Cu}_{12}\textrm{Sb}_{4}\textrm{S}_{13}$ tetrahedrite, close to $E_F$. Black dashed line marks Fermilevel. Values of $f^{'}(x)|_{E=E_{F}}$ are derivatives of DOS at Fermi level.}
  \label{fig:fig11}
\end{figure}

The bandstructures of $\textrm{Mg}_{x}\textrm{(6b)}\textrm{Cu}_{12}\textrm{Sb}_{4}\textrm{S}_{13}$ tetrahedrites are presented in a form of two highest valence and two lowest conduction bands in \textbf{Figure \ref{fig:fig9}}. Unmodified tetrahedrite, $\textrm{Cu}_{12}\textrm{Sb}_{4}\textrm{S}_{13}$ is a p--type material. As expected, introduction of magnesium, which yields 2 additional electrons, results in carrier compensation (up to $=1.0$). For $x=1.0$, tetrahedrite is a semiconductor. Increasing Mg concentration above that results in a n--type conductivity, as Fermi level is shifted into conduction band and carrier concentration increases. Band gap decreases with an increase in Mg concentration (values of band gaps are 0.98, 0.80, 0.55, 0.40 and 0.23 eV for x = 0.0, 0.5, 1.0, 1.5 and 2.0 respectively). In all cases our calculations predict indirect transition from valence band maximum at $\Gamma$ point to conduction band minimum positioned toward H point. Partial Density of States plots in \textbf{Figure \ref{fig:fig10}} (for x=1) should also help visualize electronic states of different character. Valence band maximum has character predominantly related to Cu(3d) electrons, with some minor share coming from S(3p) electrons. Conduction band minimum is composed mainly of states coming from Sb(5p), Cu(3d) and S(3p) electrons. States related to magnesium's electrons provide very minor contribution to total DOS near $E_F$.
Finally, \textbf{Figure \ref{fig:fig11}} showcases total DOS near $E_F$. Inset value of the derivative at $E_F$, i.e. slope of DOS, is proportional to Seebeck coefficient. It is worth noting that the derivative for $x=0.5$ increases significantly, compared to unmodified tetrahedrite. Change of sign in derivative implies change in sign for Seebeck coefficient as well as change from p-- to n--type conductivity.

\section{Conclusions}

In this work, first-principle calculations had been carried out to describe various structures of Cu-rich and Mg-doped tetrahedrite. Based on the results, $\square$(24g) is preferred by copper atoms in case of Cu-rich and $\square$(6b) is preferred in Mg-doped tetrahedrite. Other variations, such as magnesium occupying copper’s 12e/12d position, are also possible – showcasing ”flexibility” of the structure. However, there seems to be solubility limit of around $x=1.5$ Mg a.p.f.u. Impact of Mg/Cu excess atoms at 6b/24g Wyckoff position on local environment is different. Introduction of magnesium yields higher strain on its local surrounding than copper, despite it having slightly lower ionic radii (57 and 60 pm respectively for Mg and Cu \cite{shannon1976revised}). Structural void 6b is highly symmetrical, with four S and six Cu atoms, which increase their distance to $\square$(6b) with introduction of additional Cu/Mg. This is suitable environment for Mg atom, which is comparably an electropositive element in tetrahedrite and has tendency to form directionless, ionic, closed-shell bonds with nearby sulfur atoms. On the other hand, introduction of Cu/Mg into 24g site results in nearby Cu(2) atoms being pushed away from the center of S\textsubscript{3} triangle. The “24g” systems not only yields higher increase in local distances to $\square$(24g), but also disrupts symmetry of the structure to a higher degree. For the most cases, it is expected to slightly increase in average S\textsubscript{tri}, and as a result decrease ADP of Cu(2). Topological results show that weak Cu(2) – Sb bond indeed exist and tetrahedrite is a closed-shell, ionic crystal with some degree of covalency. With more magnesium being introduced into $\square$(6b), average bond strength decreases and whole structures becomes strained. Finally, n--type conductivity should be theoretically feasible, as indicated by band structure calculations for $x>1.0$; however it is likely unstable and might yield serious challenge in experimental synthesis. Nonetheless, our results present possibility of introduction of magnesium into tetrahedrite, with prospect of obtaining n--type conductivity. Other alkaline earth metals or alkali metals, forming directionless, closed-shell bonds might perhaps be a good choice for searching n--type tetrahedrite.
\\
\\
\textbf{Acknowledgements}
\\
This research project was supported by the program “The Excellence Initiative - Research University" and the PLGrid Infrastructure.
\\
\\
\textbf{Conflict of Interest}
\\
The authors declare no conflict of interest.
\\
\\
\textbf{Availability Statement}
\\
The data that support the findings of this study is available from the corresponding author upon reasonable request.
\\
\\
\textbf{CRediT author statement}
\\
\textbf{Krzysztof Kapera:} Conceptualization, Data Curation, Formal analysis, Investigation, Visualization, Writing - Original Draft. \textbf{Andrzej Koleżyński:} Conceptualization, Funding acquisition, Project administration, Resources, Supervision, Writing - Review and Editing.

\bibliographystyle{elsarticle-num} 
\bibliography{Bibliography}

\newpage

\appendix
\section{\textsc{Critic2} tables with topological results}
\setcounter{table}{0}

\begin{table}[H]
    \caption{Bond length \textbf{R}, eigenvalues of the hessian matrix \boldmath$\lambda_{1-3}$, electron density \boldmath$\rho(r)$, laplacian \boldmath$\nabla^2\rho(r)$, potential energy \boldmath$V[\rho(r)]$, total local energy density \boldmath$H_e[\rho(r)]$, kinetic energy density \boldmath$G[\rho(r)]$, bond’s valency \textbf{Val} and number of given bond type per unit cell \textbf{Mult} for $\textrm{Mg}_{0.0}\textrm{(6b)}\textrm{Cu}_{12}\textrm{Sb}_{4}\textrm{S}_{13}$ tetrahedrite. All values are at the Bond Critical Point (BCP).}
    \label{tab:a1}
    \begin{adjustbox}{max width=\textwidth}
    \begin{tabular}[htbp]{c l c c c c c c c c c c c}
    \hline
    \multicolumn{2}{c}{\multirow{2}{*}{{\LARGE $x=0.0$}}} & R & $\lambda_1$ & $\lambda_2$& $\lambda_3$ & $\rho(r)$ & $\nabla^2\rho(r)$ & $V[\rho(r)]$ & $H_e[\rho(r)]$ & $G[\rho(r)]$ & Val & Multi\\
    \multicolumn{2}{c}{} & {\small[{\AA}}] & {\small[au$\cdot10^{-2}$]} & {\small[au$\cdot10^{-2}$]} & {\small[au$\cdot10^{-2}$]} & {\small[au$\cdot10^{-2}$]} & {\small[au$\cdot10^{-2}$]} & {\small[au]} & {\small[au]} & {\small[au]} &  & \\
    \hline
    \multirow{1}{*}{r\textsubscript{1}} & \multirow{1}{*}{Cu(2) -- Sb} & 3.221 & -0.954	& -0.905 & 4.280 & 1.814 & 2.421 & -0.078 & -0.036 & 0.0422 & 0.167 & 24\\
    \hline
    \multirow{1}{*}{r\textsubscript{2}} & \multirow{1}{*}{S(2) -- Cu(2)} & 2.245 & -7.932 & -7.880 & 26.383 & 8.086 & 10.572 & -0.921 & -0.447 & 0.474 & 0.333 & 12\\
    \hline
    \multirow{1}{*}{r\textsubscript{3}} & \multirow{1}{*}{S(1) -- Cu(1)} & 2.259 & -7.934 & -7.885 & 27.392 & 8.126 & 11.573 & -0.839 & -0.405 & 0.434 & 0.250 & 48\\
    \hline
    \multirow{1}{*}{r\textsubscript{4}} & \multirow{1}{*}{S(1) -- Sb} & 2.468 & -7.682 & -7.582 & 20.141 & 8.263 & 4.876 & -1.028 & -0.508 & 0.520 & 0.833 & 24\\
    \hline
    \multirow{1}{*}{r\textsubscript{5}} & \multirow{1}{*}{S(1) -- Cu(2)} & 2.221 & -8.774 & -8.609 & 28.147 & 8.735 & 10.764 & -0.923 & -0.488 & 0.475 & 0.667 & 24\\
    \end{tabular}
    \end{adjustbox}
\end{table}

\begin{table}[H]
    \caption{Bond length \textbf{R}, eigenvalues of the hessian matrix \boldmath$\lambda_{1-3}$, electron density \boldmath$\rho(r)$, laplacian \boldmath$\nabla^2\rho(r)$, potential energy \boldmath$V[\rho(r)]$, total local energy density \boldmath$H_e[\rho(r)]$, kinetic energy density \boldmath$G[\rho(r)]$, bond’s valency \textbf{Val} and number of given bond type per unit cell \textbf{Mult} for $\textrm{Mg}_{1.0}\textrm{(6b)}\textrm{Cu}_{12}\textrm{Sb}_{4}\textrm{S}_{13}$ tetrahedrite. All values are at the Bond Critical Point (BCP).}
    \label{tab:a2}
    \begin{adjustbox}{max width=\textwidth}
    \begin{tabular}[htbp]{c l c c c c c c c c c c c}
    \hline
    \multicolumn{2}{c}{\multirow{2}{*}{{\LARGE $x=1.0$}}} & R & $\lambda_1$ & $\lambda_2$& $\lambda_3$ & $\rho(r)$ & $\nabla^2\rho(r)$ & $V[\rho(r)]$ & $H_e[\rho(r)]$ & $G[\rho(r)]$ & Val & Multi\\
    \multicolumn{2}{c}{} & {\small[{\AA}}] & {\small[au$\cdot10^{-2}$]} & {\small[au$\cdot10^{-2}$]} & {\small[au$\cdot10^{-2}$]} & {\small[au$\cdot10^{-2}$]} & {\small[au$\cdot10^{-2}$]} & {\small[au]} & {\small[au]} & {\small[au]} &  & \\
    \hline
    \multirow{4}{*}{r\textsubscript{1}} & \multirow{4}{*}{Cu(2) -- Sb} & 3.341 & -0.693 & -0.621 & 3.516 & 1.476 & 2.202 & -0.029 & -0.012 & 0.017 & 0.120 & 4\\
    & & 3.350 & -0.739 & -0.667 & 3.557 & 1.478 & 2.151 & -0.014 & -0.004 & 0.010 & 0.117 & 8\\
    & & 3.347 & -0.732 & -0.677 & 3.569 & 1.481 & 2.159 & -0.007 & -0.001 & 0.006 & 0.118 & 4\\
    & & 3.118 & -1.193 & -1.146 & 5.287 & 2.118 & 2.949 & -0.175 & -0.084 & 0.091 & 0.220 & 8\\
    \hline
    \multirow{4}{*}{r\textsubscript{2}} & \multirow{4}{*}{S(2) -- Cu(2)} & 2.267 & -6.057 & -5.733 & 27.316 & 7.476 & 15.526 & -0.895 & -0.428 & 0.467 & 0.314 & 2\\
    & & 2.282 & -6.032 & -5.554 & 26.636 & 7.271 & 15.050 & -0.875 & -0.419 & 0.457 & 0.301 & 4\\
    & & 2.282 & -5.950 & -5.418 & 26.366 & 7.260 & 14.998 & -0.872 & -0.417 & 0.455 & 0.301 & 4\\
    & & 2.254 & -6.375 & -5.904 & 28.214 & 7.667 & 15.935 & -0.916 & -0.438 & 0.478 & 0.325 & 2\\
    \hline
    \multirow{6}{*}{r\textsubscript{3}} & \multirow{6}{*}{S(1) -- Cu(1)} & 2.287 & -5.780 & -5.406 & 26.937 & 7.353 & 15.752 & -0.792 & -0.376 & 0.416 & 0.232 & 8\\
    & & 2.417 & -4.070 & -3.806 & 18.967 & 5.177 & 11.091 & -0.558 & -0.265 & 0.293 & 0.163 & 8\\
    & & 2.416 & -4.097 & -3.831 & 19.093 & 5.212 & 11.165 & -0.561 & -0.267 & 0.295 & 0.164 & 8\\
    & & 2.261 & -6.119 & -6.098 & 29.025 & 7.771 & 16.808 & -0.839 & -0.399 & 0.441 & 0.249 & 8\\
    & & 2.261 & -6.113 & -6.047 & 28.885 & 7.760 &  16.726 & -0.840 & -0.399 & 0.441 & 0.249 & 8\\
    & & 2.286 & -5.819 & -5.563 & 27.194 & 7.386 & 15.812 & -0.792 & -0.376 & 0.416 & 0.232 & 8\\
    \hline
    \multirow{4}{*}{r\textsubscript{4}} & \multirow{4}{*}{S(1) -- Sb} & 2.433 & -6.024 & -5.867 & 17.589 & 8.452 & 5.698 & -1.080 & -0.533 & 0.547 & 0.916 & 4\\
    & & 2.540 & -4.510 & -4.393 & 13.169 & 6.328 & 4.266 & -0.808 & -0.399 & 0.410 & 0.686 & 8\\
    & & 2.466 & -5.509 & -5.366 & 16.086 & 7.730 & 5.211 & -0.987 & -0.487 & 0.500 & 0.838 & 4\\
    & & 2.468 & -5.480 & -5.337 & 16.000 & 7.688 & 5.183 & -0.982 & -0.485 & 0.498 & 0.833 & 8\\
    \hline
    \multirow{4}{*}{r\textsubscript{5}} & \multirow{4}{*}{S(1) -- Cu(2)} & 2.244 & -6.480 & -6.209 & 29.222 & 7.984 & 16.533 & -0.889 & -0.424 & 0.465 & 0.625 & 4\\
    & & 2.281 & -5.905 & -5.506 & 27.096 & 7.495 & 15.685 & -0.844 & -0.403 & 0.442 & 0.566 & 8\\
    & & 2.232 & -6.727 & -6.474 & 30.142 & 8.215 & 16.941 & -0.906 & -0.432 & 0.474 & 0.647 & 4\\
    & & 2.239 & -6.567 & -6.346 & 29.560 & 8.116 & 16.648 & -0.898 & -0.428 & 0.470 & 0.634 & 8\\
    \hline
        \multirow{1}{*}{r\textsubscript{6}} & \multirow{1}{*}{S(1) -- Mg} & 2.350 & -3.212 & -3.069 & 23.048 & 3.855 & 16.767 & -1.008 & -0.483 & 0.525 & 0.500 & 8\\
    \end{tabular}
    \end{adjustbox}
\end{table}

\begin{table}[H]
    \caption{Bond length \textbf{R}, eigenvalues of the hessian matrix \boldmath$\lambda_{1-3}$, electron density \boldmath$\rho(r)$, laplacian \boldmath$\nabla^2\rho(r)$, potential energy \boldmath$V[\rho(r)]$, total local energy density \boldmath$H_e[\rho(r)]$, kinetic energy density \boldmath$G[\rho(r)]$, bond’s valency \textbf{Val} and number of given bond type per unit cell \textbf{Mult} for $\textrm{Mg}_{1.5}\textrm{(6b)}\textrm{Cu}_{12}\textrm{Sb}_{4}\textrm{S}_{13}$ tetrahedrite. All values are at the Bond Critical Point (BCP).}
    \label{tab:a3}
    \begin{adjustbox}{max width=\textwidth}
    \begin{tabular}[htbp]{c l c c c c c c c c c c c}
    \hline
    \multicolumn{2}{c}{\multirow{2}{*}{{\LARGE $x=1.5$}}} & R & $\lambda_1$ & $\lambda_2$& $\lambda_3$ & $\rho(r)$ & $\nabla^2\rho(r)$ & $V[\rho(r)]$ & $H_e[\rho(r)]$ & $G[\rho(r)]$ & Val & Multi\\
    \multicolumn{2}{c}{} & {\small[{\AA}}] & {\small[au$\cdot10^{-2}$]} & {\small[au$\cdot10^{-2}$]} & {\small[au$\cdot10^{-2}$]} & {\small[au$\cdot10^{-2}$]} & {\small[au$\cdot10^{-2}$]} & {\small[au]} & {\small[au]} & {\small[au]} &  & \\
    \hline
    \multirow{2}{*}{r\textsubscript{1}} & \multirow{2}{*}{Cu(2) -- Sb} & 3.328 & -0.786 & -0.708 & 3.675 & 1.533 & 2.181 & -0.011 & -0.003 & 0.008 & 0.125 &  12\\
    & & 2.973 & -1.652 & -1.651 & 6.709 & 2.726 & 3.406 & -0.277 & -0.134 & 0.143 & 0.325 & 12\\
    \hline
    \multirow{2}{*}{r\textsubscript{2}} & \multirow{2}{*}{S(2) -- Cu(2)} & 2.303 & -5.766 & -5.132 & 25.188 & 7.014 & 14.290 & -0.830 & -0.397 & 0.433 & 0.285 & 6\\
    & & 2.296 & -5.871 & -5.357 & 25.862 & 7.085 & 14.634 & -0.834 & -0.399 & 0.435 & 0.290 & 6\\
    \hline
    \multirow{2}{*}{r\textsubscript{3}} & \multirow{2}{*}{S(1) -- Cu(1)} & 2.393 & -4.439 & -3.747 & 20.928 & 6.165 & 12.743 & -0.634 & -0.301 & 0.333 & 0.174 & 24\\
    & & 2.261 & -6.114 & -6.054 & 28.837 & 7.764 & 16.669 & -0.828 & -0.393 & 0.435 & 0.249 & 24\\
    \hline
    \multirow{2}{*}{r\textsubscript{4}} & \multirow{2}{*}{S(1) -- Sb} & 2.528 & -5.658 & -5.455 & 14.894 & 6.544 & 3.781 & -0.863 & -0.427 & 0.436 & 0.718 & 12\\
    & & 2.473 & -6.462 & -5.968 & 17.632 & 7.372 & 4.186 & -0.892 & -0.441 & 0.451 & 0.823 & 12\\
    \hline
    \multirow{2}{*}{r\textsubscript{5}} & \multirow{2}{*}{S(1) -- Cu(2)} & 2.291 & -5.750 & -5.360 & 26.170 & 7.346 & 15.060 & -0.808 & -0.385 & 0.423 & 0.551 & 12\\
    & & 2.236 & -6.605 & -6.413 & 29.625 & 8.162 & 16.607 & -0.895 & -0.427 & 0.468 & 0.639 & 12\\
    \hline
        \multirow{1}{*}{r\textsubscript{6}} & \multirow{1}{*}{S(1) -- Mg} & 2.354 & -3.193 & -2.974 & 22.866 & 3.875 & 16.700 & -0.975 & -0.467 & 0.508 & 0.500 & 12\\
    \end{tabular}
    \end{adjustbox}
\end{table}

\begin{table}[H]
    \caption{Bond length \textbf{R}, eigenvalues of the hessian matrix \boldmath$\lambda_{1-3}$, electron density \boldmath$\rho(r)$, laplacian \boldmath$\nabla^2\rho(r)$, potential energy \boldmath$V[\rho(r)]$, total local energy density \boldmath$H_e[\rho(r)]$, kinetic energy density \boldmath$G[\rho(r)]$, bond’s valency \textbf{Val} and number of given bond type per unit cell \textbf{Mult} for $\textrm{Mg}_{2.0}\textrm{(6b)}\textrm{Cu}_{12}\textrm{Sb}_{4}\textrm{S}_{13}$ tetrahedrite. All values are at the Bond Critical Point (BCP).}
    \label{tab:a4}
    \begin{adjustbox}{max width=\textwidth}
    \begin{tabular}[htbp]{c l c c c c c c c c c c c}
    \hline
    \multicolumn{2}{c}{\multirow{2}{*}{{\LARGE $x=2.0$}}} & R & $\lambda_1$ & $\lambda_2$& $\lambda_3$ & $\rho(r)$ & $\nabla^2\rho(r)$ & $V[\rho(r)]$ & $H_e[\rho(r)]$ & $G[\rho(r)]$ & Val & Multi\\
    \multicolumn{2}{c}{} & {\small[{\AA}}] & {\small[au$\cdot10^{-2}$]} & {\small[au$\cdot10^{-2}$]} & {\small[au$\cdot10^{-2}$]} & {\small[au$\cdot10^{-2}$]} & {\small[au$\cdot10^{-2}$]} & {\small[au]} & {\small[au]} & {\small[au]} &  & \\
    \hline
    \multirow{4}{*}{r\textsubscript{1}} & \multirow{4}{*}{Cu(2) -- Sb} & 3.278 & -0.857 & -0.776 & 3.951 & 1.642 & 2.319 & -0.039 & -0.017 & 0.022 & 0.143 & 8\\
    & & 3.050 & -1.482 & -1.390 & 5.989 & 2.441 & 3.117 & -0.194 & -0.093 & 0.101 & 0.264 & 4\\
    & & 2.901 & -1.993 & -1.981 & 7.488 & 3.112 & 3.515 & -0.338 & -0.164 & 0.173 & 0.395 & 8\\
    & & 2.895 & -2.074 & -2.038 & 7.628 & 3.195 & 3.517 & -0.340 & -0.166 & 0.175 & 0.402 & 4\\
    \hline
    \multirow{4}{*}{r\textsubscript{2}} & \multirow{4}{*}{S(2) -- Cu(2)} & 2.333 & -5.392 & -4.807 & 23.683 & 6.604 & 13.484 & -0.771 & -0.369 & 0.402 & 0.262 & 2\\
    & & 2.345 & -5.256 & -4.628 & 22.906 & 6.477 & 13.022 & -0.764 & -0.366 & 0.398 & 0.254 & 2\\
    & & 2.331 & -5.452 & -4.771 & 23.621 & 6.670 & 13.397 & -0.780 & -0.373 & 0.407 & 0.264 & 4\\
    & & 2.316 & -5.639 & -5.095 & 24.756 & 6.836 & 14.022 & -0.797 & -0.381 & 0.416 & 0.275 & 4\\
    \hline
    \multirow{6}{*}{r\textsubscript{3}} & \multirow{6}{*}{S(1) -- Cu(1)} & 22.388 & -4.480 & -3.856 & 21.067 & 6.228 & 12.731 & -0.634 & -0.301 & 0.333 & 0.177 & 8\\
    & & 2.360 & -4.734 & -4.187 & 22.530 & 6.582 & 13.609 & -0.684 & -0.325 & 0.359 & 0.190 & 8\\
    & & 2.257 & -6.155 & -6.093 & 29.068 & 7.807 & 16.821 & -0.831 & -0.394 & 0.436 & 0.251 & 8\\
    & & 2.383 &	-4.537 & -3.896 & 21.391 & 6.285 & 12.958 & -0.641 & -0.304 & 0.337 & 0.179 & 8\\
    & & 2.354 & -4.796 & -4.272 & 22.956 & 6.635 & 13.888 & -0.700 & -0.333 & 0.367 & 0.194 & 8\\
    & & 2.257 & -6.201 & -6.109 & 28.928 & 7.848 & 16.617 & -0.834 & -0.396 & 0.438 & 0.252 & 8\\
    \hline
    \multirow{4}{*}{r\textsubscript{4}} & \multirow{4}{*}{S(1) -- Sb} & 2.639 & -4.968 & -4.769 & 13.263 & 5.919 & 3.525 & -0.755 & -0.373 & 0.382 & 0.526 & 8\\
    & & 2.599 & -5.298 & -5.273 & 14.024 & 6.288 & 3.454 & -0.849 & -0.420 & 0.429 & 0.586 & 4\\
    & & 2.631 & -5.057 & -4.891 & 13.408 & 6.024 & 3.461 & -0.759 & -0.375 & 0.384 & 0.536 & 4\\
    & & 2.473 & -6.325 & -6.843 & 16.401 & 7.440 & 3.233 & -1.142 & -0.567 & 0.575 & 0.822 & 8\\
    \hline
    \multirow{4}{*}{r\textsubscript{5}} & \multirow{4}{*}{S(1) -- Cu(2)} & 2.313 & -5.383 & -5.081 & 24.636 & 7.066 & 14.171 & -0.765 & -0.365 & 0.400 & 0.519 & 4\\
    & & 2.306 & -5.472 & -5.199 & 25.317 & 7.185 & 14.645 & -0.789 & -0.376 & 0.413 & 0.529 & 4\\
    & & 2.249 & -6.345 & -6.192 & 28.656 & 7.932 & 16.119 & -0.872 & -0.416 & 0.456 & 0.618 & 8\\
    & & 2.302 & -5.596 & -5.245 & 25.426 & 7.217 & 14.585 & -0.784 & -0.374 & 0.410 & 0.536 & 8\\
    \hline
        \multirow{3}{*}{r\textsubscript{6}} & \multirow{3}{*}{S(1) -- Mg} & 2.381 & -2.953 & -2.715 & 21.292 & 3.715 & 15.625 & -0.929 & -0.445 & 0.484 & 0.500 & 4\\
        & & 2.378 & -2.914 & -2.567 & 21.240 & 3.716 & 15.760 & -0.948 & -0.454 & 0.494 & 0.500 & 4\\
        & & 2.373 & -3.039 & -2.710 & 21.777 & 3.789 & 16.029 & -0.940 & -0.450 & 0.490 & 0.500 & 8\\
    \end{tabular}
    \end{adjustbox}
\end{table}

\end{document}